\documentclass[]{scrartcl}
\usepackage{graphics}
\usepackage[english]{babel}
\usepackage{amsmath} 
\usepackage{amsfonts}
\usepackage{latexsym}
\usepackage{cuted}
\usepackage{graphicx}
\usepackage{graphics}
\usepackage{float}
\usepackage{breqn}
\usepackage{amssymb}
\usepackage{authblk}
\usepackage{multirow}
\usepackage{makeidx}
\usepackage{dblfloatfix}
\usepackage[section]{placeins}

\PassOptionsToPackage{hyphens}{url}
\usepackage{hyperref}
\usepackage{url}
\usepackage{graphics}
\usepackage[english]{babel}
\usepackage{amsmath} 
\usepackage{amsfonts}
\usepackage{latexsym}
\usepackage{graphicx}
\usepackage{graphics}
\usepackage{float}
\usepackage{hyperref}
\usepackage{amssymb}
\usepackage{authblk}
\usepackage{multirow}
\usepackage{makeidx}
\usepackage{dblfloatfix}
\usepackage[section]{placeins}
\usepackage{url}
\usepackage{flushend}
\begin{document}

\title{An Improved Test of the General Relativistic Effect of Frame-Dragging  Using the LARES and LAGEOS Satellites}
\author[1,2]{Ignazio Ciufolini\thanks{ignazio.ciufolini@unisalento.it}}
\author[3]{Antonio Paolozzi}
\author[4]{Erricos C. Pavlis}
\author[3]{Giampiero Sindoni}
\author[5]{John Ries}
\author[6]{Richard Matzner}
\author[7]{Rolf Koenig}
\author[2,3]{Claudio Paris}

\author[8]{Vahe Gurzadyan}
\author[9]{ Roger Penrose}

\affil[1]{\footnotesize Dip. Ingegneria dell'Innovazione, Universit\`a del Salento, Lecce, Italy}
\affil[2]{Museo della fisica e Centro studi e ricerche Enrico Fermi, Rome, Italy}
\affil[3]{Scuola di Ingegneria Aerospaziale, Sapienza Universit\`a di Roma, Italy}
\affil[4]{Joint Center for Earth Systems Technology (JCET), University of Maryland, Baltimore County, USA}

\affil[5]{Center for Space Research, University of Texas at Austin, Austin, USA}
\affil[6]{Theory Center, University of Texas at Austin, Austin, USA}

\affil[7]{Helmholtz Centre Potsdam, GFZ German Research Centre for Geosciences, Potsdam, Germany}

\affil[8]{Center for Cosmology and Astrophysics, Alikhanian National Laboratory and Yerevan State University, Yerevan, Armenia}

\affil[9]{Mathematical Institute, University of Oxford, Oxford, UK}

\renewcommand\Authands{ and }

\date{}
\maketitle
%

%
%
%
%
%

%

\abstract{We report the improved test of frame-dragging, an intriguing phenomenon predicted by Einstein's General Relativity, obtained using 7 years of Satellite Laser Ranging (SLR) data of the satellite LARES (ASI, 2012) and 26 years of SLR data of LAGEOS (NASA, 1976) and LAGEOS 2 (ASI and NASA, 1992). We used static part and temporal variations of the Earth gravity field obtained by the space geodesy mission GRACE (NASA and DLR) and in particular the static Earth's gravity field model GGM05S with its 7-day variations of the lowest degree Earth's spherical harmonics. We used the orbital estimator GEODYN (NASA). We measured frame-dragging to be equal to $(0.9910 \pm 0.0006) \pm 0.02 \; to \, \pm 0.04$, where 1 is the theoretical prediction of General Relativity normalized to its frame-dragging value, 0.0006 is the formal 1-sigma statistical error of our fit and $\pm 0.02 \; to \, \pm 0.04$ is the estimated systematic error due to the errors in the Earth's gravity field determination. The 2\% uncertainty is obtained by propagating the GGM05S systematic errors in the observables, plus other smaller error sources, whereas 4\% is a conservative uncertainty obtained by multiplying the systematic errors of GGM05S by a factor three. }

%

%

\section{General Relativity, Dragging of Inertial Frames and the Objectives of the LARES Space Mission}
\label{sec:1}

Einstein's gravitational theory of General Relativity is fundamental to understand our universe \cite{mtw,wei,novzel,ciuwhe}. It has a number of outstanding experimental verifications \cite{will,ciuwhe,tur},
among which are the recent impressive LIGO laser interferometers direct detections of gravitational waves and observation of black holes, and of their collision, through the emission of gravitational waves \cite{ligo1,ligo2}.

LARES (LAser RElativity Satellite) \cite{ciuetal11} is a laser-ranged satellite of ASI, the Italian Space Agency, dedicated to test General Relativity and 
fundamental physics, and to measurements of space geodesy and geodynamics. 
Among the tests of General Relativity, the main objective of LARES is a measurement of dragging of inertial frames, or frame-dragging, with an accuracy 
of a few percent. In addition to the test of frame-dragging, LARES, together with the LAGEOS (LAser GEOdynamcs Satellite of NASA) \cite{lag} and LAGEOS 2 (of ASI and NASA), has recently provided a test of the weak equivalence principle \cite{EPpaper}, at the foundations of General Relativity and other viable
gravitational theories, with an accuracy of about $10^{-9}$, at a previously untested range between about 7820 km and 12270 km, and using previously 
untested materials of a tungsten alloy (the material of LARES) and aluminum-brass (the material of LAGEOS and LAGEOS 2).
The orbital parameters and characteristics of the LARES and LAGEOS satellites are provided in the next section.

Frame-dragging \cite{ein} is an intriguing phenomenon of General Relativity: in Einstein's gravitational theory the inertial frames, which can only be defined 
locally (according to the equivalence principle \cite{mtw,wei,ciuwhe}), have no fixed direction with respect to the distant stars but are instead 
dragged by the currents of mass-energy such as the rotation of a body, e.g., the rotation of the Earth (the axes of the local inertial frames are determined in General Relativity by local test-gyroscopes.) For a detailed description of such intriguing phenomenon and its fascinating astrophysical implications around spinning black holes, we refer to \cite{tho,mtw,ciuwhe,ciunat07,ocon,kope2}. 
Here we just observe that the 2015-2018 detection of gravitational waves by the LIGO detectors \cite{ligo1,ligo2} are also based on computer simulations
of the collision of spinning black holes and spinning neutron stars to form a spinning black hole. In such astrophysical processes, frame-dragging 
plays a key role \cite{thorne}.

The effect of frame-dragging on the orbit of a satellite is called the Lense-Thirring effect \cite{len}. The orbital plane of a satellite may be thought of as a gyroscope (it would be a perfect gyroscope if the satellite were orbiting under the influence of a central gravitational force only, due to a spherically symmetric body). 
Indeed, the line of the nodes (intersection of the satellite's orbital plane with the equatorial plane of the central body \cite{kau,kop}) has a frame-dragging shift described by the
Lense-Thirring effect.

Between 2004 and 2016, frame-dragging was tested in a series of papers with an accuracy between about 10\% and 5\% using the two LAGEOS 
satellites  \cite{ciunat,rie08,rie09,ciuetal10,koe12} and the LARES and LAGEOS satellites \cite{ciuetal16,ciuetal11}.
In 2011, frame-dragging was tested using the Gravity Probe B experiment with an accuracy of about 20\% \cite{GPB}. It may be noted that GP-B measured the spin-spin effect due to frame-dragging, whereas
 LARES measured the spin-orbit frame-dragging effect. Thus, for example, gravitational theories with torsion (antisymmetric connection) may imply a different outcome for these two frame-dragging effects (for gravitational theories predicting frame-dragging effects different from General Relativity see, e.g., \cite{smith}). Furthermore the errors in each test have no commonality with each other, thus these tests complement each other and provide a more robust overall test of General Relativity.

\section{The LARES, LAGEOS, LAGEOS 2 and GRACE satellites and the method of analysis}
\label{sec:2}

The LARES satellite \cite{pao1} of ASI was successfully launched on 13th February 2012 using the VEGA launch vehicle (of ESA-ASI-ELV-AVIO). It has 
a semimajor axis of 7821 km, orbital eccentricity of 0.0008 and inclination of $69.5^\circ$. Its mass is 386.8 kg, its diameter 36.4 cm and is covered
with 92 retro-reflectors (CCRs) reflecting back the laser pulses to precisely determine the range between the satellite and the laser-ranging stations on Earth
with an uncertainty, for the best SLR (Satellite Laser Ranging) stations, of less than one centimeter \cite{ilrs}. The LAGEOS (NASA) \cite{lag} and LAGEOS 2 (ASI-NASA) satellites were respectively launched in 1976 and 1992.
They are almost identical with a diameter of 60 cm and a mass of respectively 405.38 and 432 kg. They are each covered with 426 CCRs.
LAGEOS has a semimajor axis of 12270 km, orbital eccentricity of 0.0045 and inclination of $109.84^\circ$, whereas LAGEOS 2 
a semimajor axis of 12163 km, orbital eccentricity of 0.0135 and inclination of $52.64^\circ$. 
The orbit of these three satellites can be determined with centimeter accuracy \cite{ilrs} using their observations by the International Laser Ranging Service (ILRS) \cite{ilrs} (their normal points have submillimeter accuracy) and orbital estimators such as GEODYN (NASA) \cite{geo}, UTOPIA (CSR-UT) and EPOSOC (GFZ).

The twin GRACE (Gravity Recovery and Climate Experiment) satellites of NASA and DLR (Deutsche Forschunganstalt f\"ur Luft und Raumfahrt: German Aerospace Center) were launched in 2002.
They have been orbiting at about 220 km apart, with a semimajor axis of 6856 km, orbital eccentricity of 0.005 and inclination of $89^\circ$. 
GRACE, by accurately measuring the variations of the distance between its twin satellites using GPS and a microwave ranging system, 
allowed an unprecedented accuracy in the determination of the Earth
gravity field and of its variations \cite{grace1}. The GRACE science mission ended in October 2017.
The GRACE Follow-On (GRACE-FO) space mission of NASA and GFZ (German Research Centre for Geosciences) \cite{GRACEFON} is the continuation of the GRACE mission. It was successfully launched in May 22, 2018. It will also test a new technology to dramatically improve the remarkable precision of its measurement system.

The main uncertainty in the measurement of the frame-dragging shift of the nodal line of an Earth satellite (the Lense-Thirring effect) is 
due to the uncertainty in the modelling of its nodal shift due to the deviations of the Earth gravity field from that of a perfect spherical body \cite{kau} and in particular due to the even zonal harmonics of the spherical harmonic expansion of the Earth gravity field. The even zonal harmonics describe the deviation from sphericity that are symmetric with respect to the Earth 
rotation axis and its equatorial plane; in fact the even zonal harmonics produce secular effects that cannot be separated from the secular Lense-Thirring effect using a single satellite.
Thus, the largest uncertainties in modelling the nodal shift of an Earth satellite are those due to the Earth quadrupole moment, $J_2$, and to the next higher degree
even zonal harmonic, $J_4$. Indeed, thanks to the GRACE determinations of the Earth gravity field and of its variations, the error in modeling the even zonal 
harmonics of degree strictly higher than four is at the level of a few percent only (see section 3) \cite{ciuetal10,ciuetal11,rie08,ciuetalcqg,ciuetal16}.

The method to use $n$ observables, i.e., the nodal lines of $n$ laser-ranged satellites to eliminate the uncertainties in the first $n - 1$ even zonal harmonics and to measure the Lense-Thirring effect,
was proposed in \cite{ciu89}, described in detail in 
\cite{ciuncc} and reported in \cite{ciunat,rie08,ciuetal10} to measure the Lense-Thirring effect, first using two satellites, LAGEOS and LAGEOS 2, and then using three satellites; LARES, LAGEOS and LAGEOS 2 \cite{ciuetal11}.

In the analysis reported in the present paper, for the three unknowns given by the uncertainties in $J_2$ and $J_4$ and by the Lense-Thirring effect, we need
three observables provided by the measured rates of the three nodal lines of LARES, LAGEOS and LAGEOS 2.

The combination that allows such measurement is \cite{ciuetal11,ciuetal16}:

\begin{displaymath}
\delta \, \dot \Omega^{LAGEOS \, I} + k_1 \, \delta \,
\dot \Omega^{LAGEOS \, 2}  + k_2  \delta \,
\dot \Omega^{LARES \,}  = 
\end{displaymath}

\begin{multline}
= \, \mu \, ( \, 30.68 \, + k_1 \, 31.50 \, + k_2 \, 118.50)  \, mas/yr  +\\ other \, errors \cong \mu \, (50.18  \, mas/yr)
\end{multline}

\noindent where $\delta \, \dot \Omega$ are the residuals (observed minus calculated) of the nodal rates of LAGEOS, LAGEOS 2 and LARES, determined with GEODYN 
using their SLR data, the GRACE gravity field model GGM05S \cite{rie} together with a model of the time variation, and other up to date orbital perturbations, including tides (GOT4.10 tidal model \cite{GOT}) and non-gravitational orbital perturbations effects such as solar radiation pressure. \cite{ciuetal10,ciuetal11}.

In Eq. (1) $30.68  \, mas/yr$, $31.50 \, mas/yr$ and $118.50\, mas/yr$ are the Lense-Thirring nodal rates, respectively of LAGEOS, LAGEOS 2 and LARES, predicted by General Relativity. $k_1$  and $k_2$ are the coefficients to eliminate the uncertainties in $J_2$ and $J_4$ in measuring the Lense-Thirring effect, calculated by solving the three equations of the residual nodal rates of the three satellites. They are equal to unity for LAGEOS, and on average equal to $k_1 = 0.3448$ for LAGEOS 2, and $k_2 = 0.07291$ for LARES.

A source of error is due to the time-variations of the lowest harmonics of the Earth gravity field, both their rates and long-periodic variations, and their 
lunar and solar tidal changes. However, in the present analysis, in order to be able to reach an accuracy in the test of frame-dragging of about 2\%, we have 
applied the following techniques:

(1) we determined and used the precise values of $k_1$ and $k_2$  over $each$ 15-day arc by using the precise but variable orbital elements of the three satellite's determined by 
GEODYN using the SLR data of the three satellites;

(2) over about one half of the period of our analysis, corresponding to the GRACE science phase, we applied the 7-day variations of the lowest Earth harmonics, 
compatible with the GGM05S model, directly in the orbital estimator GEODYN. Over the remaining period we applied the secular rates of the lowest harmonics determined 
by GRACE;

(3) finally, we applied the following new method to eliminate the uncertainties due to the main tidal perturbations.
First, we observe that we have about 26 years of SLR data of both LAGEOS and LAGEOS 2, whereas we only have about 7 years of SLR data of LARES, since its launch date in 2012. Furthermore,
the main tidal uncertainties in the combination of the residuals of the nodal rates of the three satellites, Eq. (1), are those affecting the nodes of LAGEOS and LAGEOS 2. The coefficient of the LARES nodal shift, $k_2 = 0.07291$, is much smaller than the coefficients of LAGEOS (unity) and LAGEOS 2 $k_1 = 0.3448$. However, the largest tidal signals of LAGEOS and LAGEOS 2 are due to the K1 tide and have a period of about 1051 days and 571 days respectively (i.e., they correspond to the well measured periods of the nodes of LAGEOS and LAGEOS 2 respectively). These long-period residual tidal perturbations can well be fitted for using the 26 years of SLR data of LAGEOS and LAGEOS 2.
Therefore, using GEODYN, we processed 26 years of SLR data of LAGEOS and LAGEOS 2 to fit for the residual amplitudes of these two main 
tidal signals of the nodes of LAGEOS and LAGEOS 2, and then removed these two residual tidal amplitudes from their orbital residuals. We then combined the node residuals of LAGEOS, LAGEOS 2 and LARES, according to Eq. (1), over the last 7 years of their SLR data and, by a Fourier analysis of the residuals of the combination, Eq. (1), we found three main periodical signals affecting their combination
(corresponding to tidal effects with well known frequencies due to the lunar and solar perturbations \cite{gurz}). One of these residual tidal signals corresponds to the K1 tide of LARES with its short node period of about 211 days, well fitted for over 7 years of LARES SLR data. We then removed these three residual tidal effects from the combination of the orbital residuals of the three satellites and fitted for a secular trend. 

After using this technique the combination of nodal rates of Eq. (1) became substantially unaffected by the nodal uncertainties due to the tidal errors, apart from smaller tidal signals with periods much shorter than the observation period of 7 years.  Furthermore, the other variations in the Earth gravity field were well modelled using the GRACE data. In the next section, we show the Fourier analysis of the residuals of the combination, Eq. (1), before and after removal of the five main residual tidal signals (Fig. 3 and Fig 4). In Fig. (5) we also present the distribution of the combination of the residuals after removing the five largest tidal signals, which shows the behaviour of a Gaussian normal distribution.

\section{Results of the measurement of frame-dragging with LARES, LAGEOS and LAGEOS 2}
\label{sec:3}

By fitting the combination of the residuals of LARES, LAGEOS and LAGEOS 2, Eq. (1), after removing the five main tidal residual signals of LAGEOS, LAGEOS 2 and LARES (using the technique explained in the previous section 2), we found a value of frame-dragging equal to 0.9910, the corresponding 
fit of the residuals is shown in Fig.1. In Fig.2 we show the cumulative combined residuals and their fit before removing the five main tidal residual signals of LAGEOS, LAGEOS 2 and LARES. The final result for the frame-dragging effect shown in Fig.1 is then: 

\begin{equation}
\mu = 0.9910 \pm 0.02
\end{equation}

Here $\mu$ = 1 is the value of frame-dragging normalized to its General Relativity value and 0.02 is the estimated total systematic error. This total systematic error can be estimated to be within about 2\% and 3\%, whether or not one multiplies the published systematic errors of the static GGM05S gravity field for a safety factor of 2, i.e., depending on the estimation of the systematic errors of the gravity field model GGM05S.
Indeed, by propagating the calibrated (including the published systematic uncertainties in their values) errors of GGM05S in the combination, Eq. (1), of the nodes, we found an error of 1.3\% of the Lense-Thirring effect. Thus, by including other smaller errors due to the other gravitational and non-gravitational perturbations (see \cite{ciuetal10,ciuetal11}), the RSS (Root Sum Squared) error is at the level of about 2\%. However, although the uncertainties for GGM05S have been carefully calibrated, this error calibration is statistical in nature. The errors are calibrated as a function of degree but cannot be expected to represent the error in any particular coefficient. To be more conservative, it is sensible to increase the uncertainty estimate of the static GGM05S gravity field by a factor of 2. In such a case, by multiplying for a factor 2 the calibrated errors of the static gravity field GGM05S, we get a total RSS error of about 3\%. The formal 1-sigma uncertainty of the fit of Fig. (1) is less than 0.001.

\begin{figure}
 \includegraphics[width=0.84\textwidth]{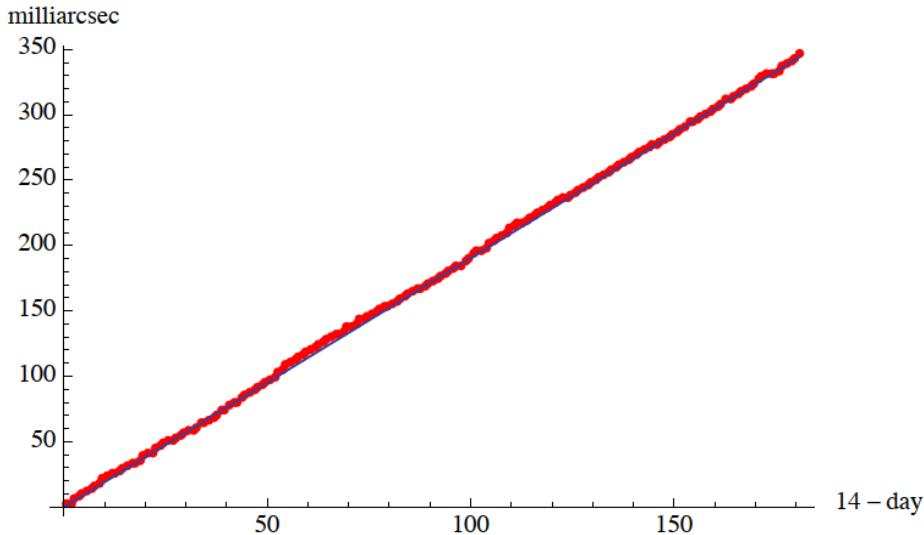}
\caption{Cumulative combined residuals of LARES, LAGEOS and LAGEOS 2 (shown in red), over about 7 years of orbital observations, after the removal the five main residual tidal signals. The solid black line is the fitted secular trend. The formal error of the fit is less than 0.001.}
\label{fig:2}       
\end{figure}

\begin{figure}
 \includegraphics[width=0.84\textwidth]{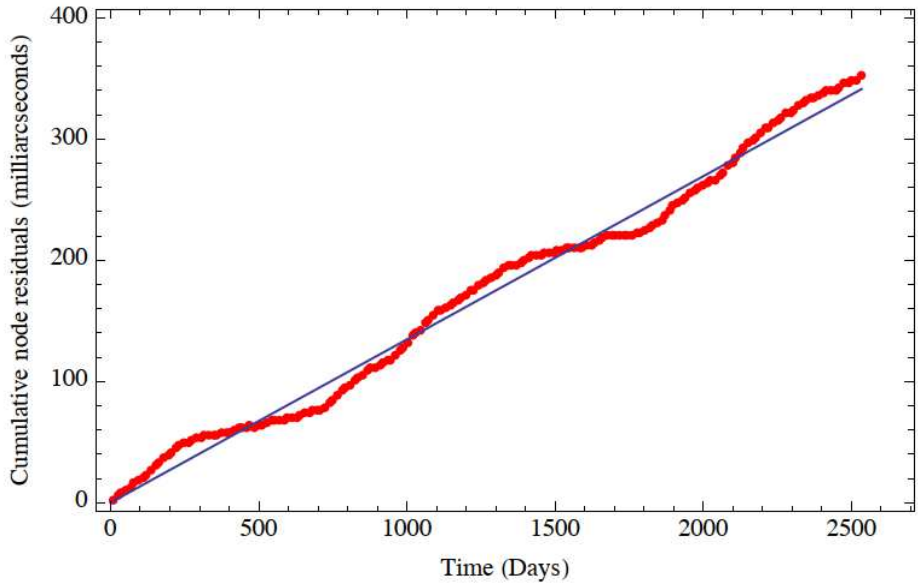}
\caption{Cumulative combined residuals of LARES, LAGEOS and LAGEOS 2 (shown in red), over about 7 years of orbital observations, fitted with a constant trend (shown with a solid black line).}
\label{fig:2}       
\end{figure}

In Fig. 3, we report the Fourier analysis of the combined residuals before removal of the five  main tidal signals on the nodes of LAGEOS, LAGEOS 2 and LARES, and in Fig. 4, the Fourier
analysis after successful removal of these five main tidal signals.
Fig. 4 shows that the main long-period tidal signals (with $n \lesssim 35$) were removed from the combination of the nodal residuals. The remaining higher frequencies do not significantly contribute to the secular trend fitted for over the observational period of 7 years.
Fig. 5 shows that the distribution of the residuals, after removal of the five main tidal signals, well approximates a normal Gaussian distribution. 

\begin{figure}
 \includegraphics[width=0.84\textwidth]{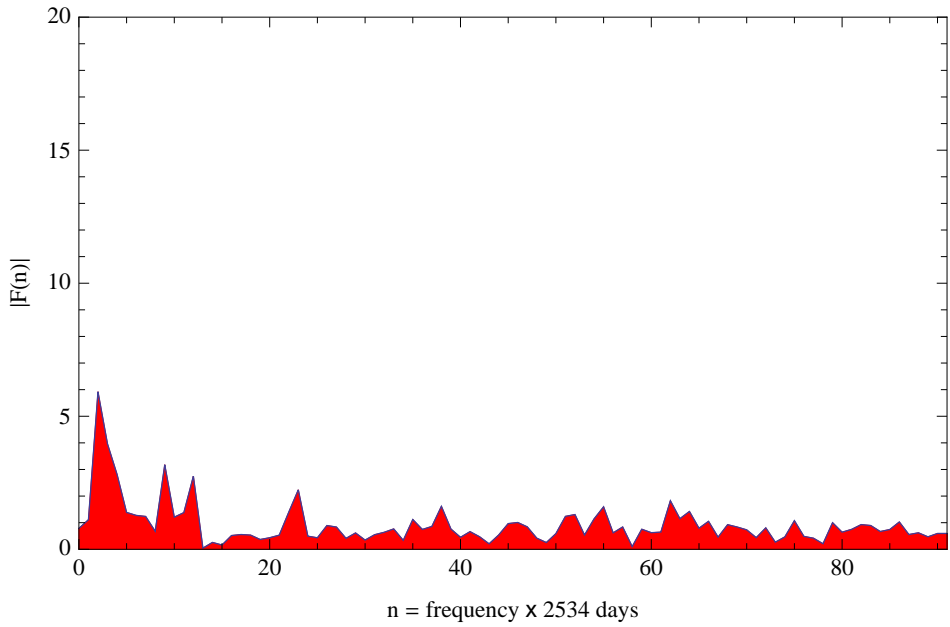}
\caption{Fourier analysis of the combined residuals of LARES, LAGEOS and LAGEOS 2 removing a constant trend, over about 7 years of orbital observations. In the horizontal axis is the frequency times 2534 days. In the vertical axis is the absolute value of the Discrete Fourier Transfom $\mid F(n) \mid$.}
\label{fig:2}       
\end{figure}

\begin{figure}
 \includegraphics[width=0.84\textwidth]{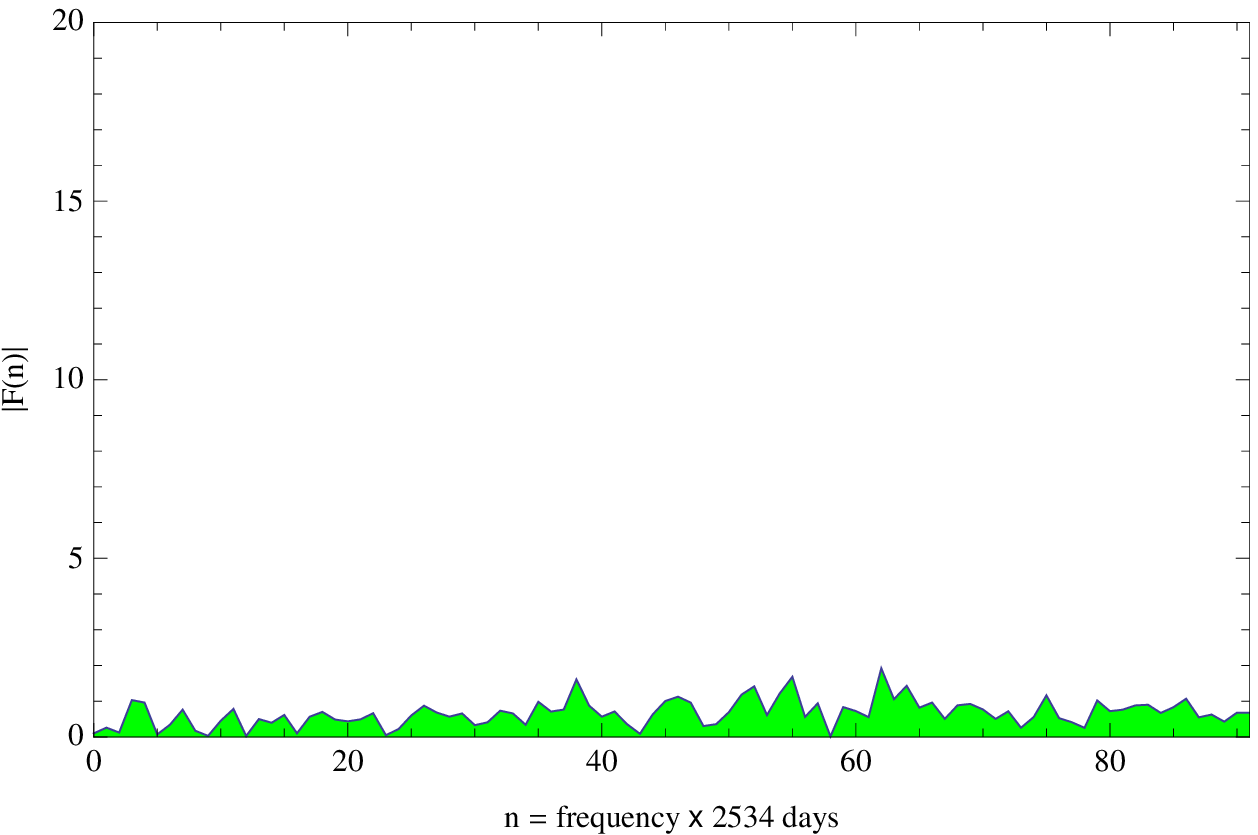}
\caption{Fourier analysis of the combined residuals of LARES, LAGEOS and LAGEOS 2, over about 7 years of orbital observations, after removal of the five main residual tidal signals and of a constant trend. In the horizontal axis is the frequency times 2534 days. In the vertical axis is the absolute value of the Discrete Fourier Transfom $\mid F(n) \mid$. The figure shows that there are no relevant {\itshape long-period} signals left after the removal of the six main tidal signals. The remaining higher frequencies $n$ do not significantly contribute to the secular trend fitted for over the observational period of 7 years.
}
\label{fig:2}       
\end{figure}

We finally performed another relevant test to answer the following question: over 
which observational period (if any) does the measurement of frame-dragging with LARES, LAGEOS and LAGEOS 2 converge to within our errors to a final result? In other words over which observational period does the test of frame-dragging with LARES, LAGEOS and LAGEOS 2 becomes a stable test, substantially independent (i.e., within the total systematic error) in the increase of the period of data analysis?
To answer this question we carried out the analysis and subsequent frame-dragging test
over successive observational periods of time, each one increased by one residual (calculated over 14 days) over the previous period.
The result is shown in Fig. 6. This figure clearly shows that after about three years of SLR observations, the measurement of frame-dragging is stable and clearly tends to its General Relativity value of 1 (normalized to its General Relativity value) well within a total systematic error of 2\%.

\begin{figure}
 \includegraphics[width=0.84\textwidth]{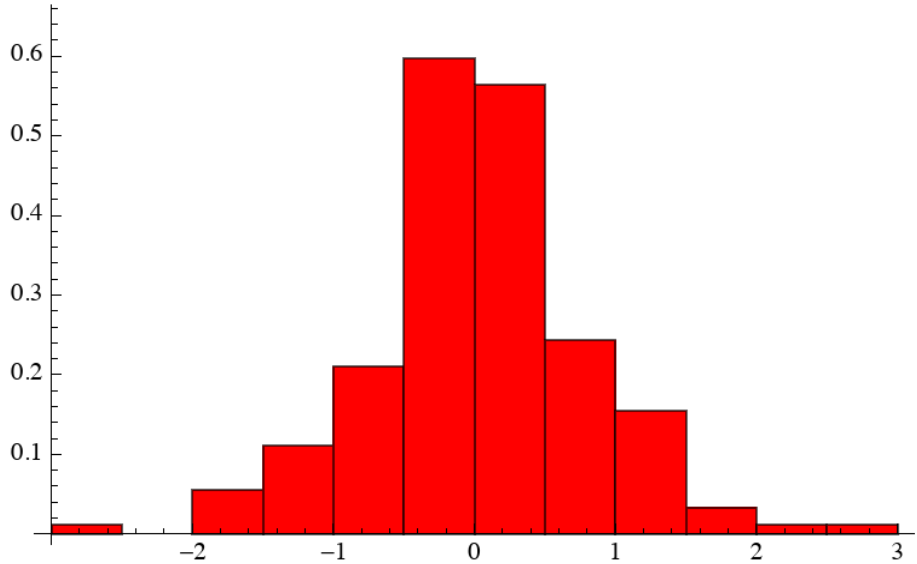}
\caption{Distribution of the post-fit combined residuals of LARES, LAGEOS and LAGEOS 2, over about 7 years of orbital observations. In the horizontal axis is the difference in milliarcsec of the residual of each arc, from the mean of the combined residuals. In the vertical axis is number of such combined residuals falling into each 0.5 milliarcsec bin. Here we have removed from the combined residuals the five main tidal signals (and the mean of the combined residuals). The distribution well approximates a Gaussian normal distribution, and supports the estimate of 2\% accuracy determined by propagating the uncertainties of GGM05S combined with other smaller errors.}
\label{fig:2}       
\end{figure}

\begin{figure}
 \includegraphics[width=0.84\textwidth]{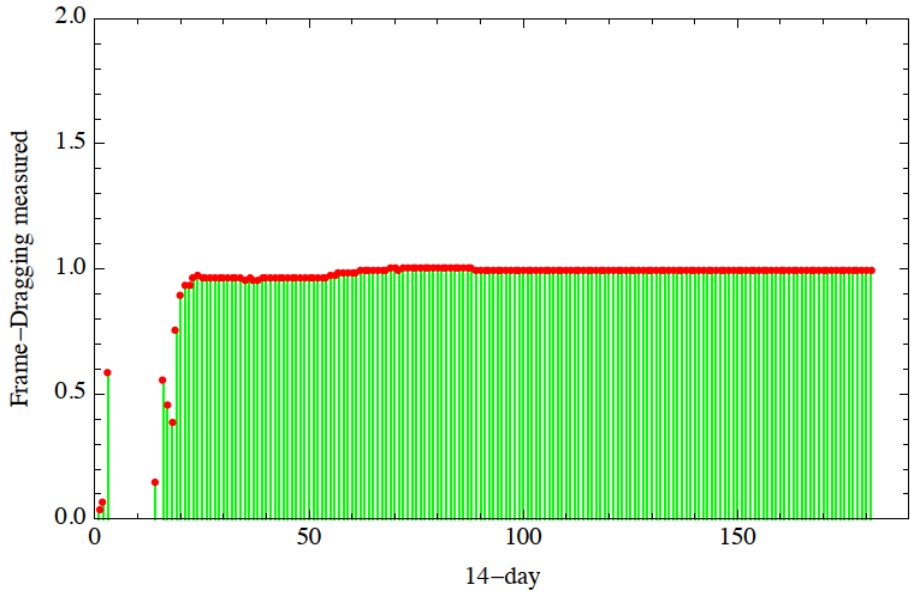}
\caption{Measurement of the Lense-Thirring effect using LARES, LAGEOS and LAGEOS 2 over increasing observational periods, i.e. each point of the curve was obtained by increasing the observational period of the previous point by 14 days, thus by increasing
the total number of residuals by one with respect to the previous point. The curve clearly tends to its General Relativity value of 1 (normalized) well within the total systematic error of 2\%.}
\label{fig:2}       
\end{figure}

\section{Conclusion}
\label{sec:4}

We analyzed 7 years of SLR data of the satellite LARES (ASI), and 26 years of SLR data of LAGEOS (NASA) and LAGEOS 2 (ASI and NASA) using the Earth gravity field model GGM05S, together with the 7-day gravity field variations obtained by the space mission GRACE (NASA and DLR). Our analysis was performed with the NASA orbital estimator GEODYN. 

By combining the residuals of these three laser-ranged satellites, we finally obtained a precise measurement of frame-dragging or Lense-Thirring effect.
Our test fully confirms the prediction of Einstein's General Relativity: we obtained a measured value of frame-dragging of 0.9910 (to be compared to its General Relativity normalized value of 1), well within the total systematic error of 2\%. Indeed, the size of the total systematic error depends mainly on the considered size of the calibrated uncertainties in the even zonal harmonics of the Earth gravity field above degree 4. A total RSS error of about 2\% corresponds to the propagation of the calibrated errors of GGM05S into the combination of the nodes of the three satellites (i.e., 1.3\% due to the error in the even zonals plus other much smaller error sources).

\section{Acknowledgements}

We gratefully acknowledge the support of the Italian Space Agency, grants  I/034/12/0, I/034/12/1 and 2015-021-R.0 and
the International Laser Ranging Service for providing high-quality laser ranging tracking of the LARES satellites.
E.C. Pavlis acknowledges the support of NASA Grants NNX09AU86G and NNX14AN50G. R. Matzner acknowledges NASA  Grant NNX09AU86G and 
J.C. Ries the support of NASA Contract NNG17VI05C.

\section{Data Availability Statement}

The SLR data of LAGEOS, LAGEOS 2 and LARES are available at: 

\noindent https://ilrs.cddis.eosdis.nasa.gov/data\_and\_products/data

\noindent /npt/index.html

%
%
%
%

\end{document}